\documentstyle[aas2pp4]{article}

\lefthead{Hirotani}
\righthead{Collimation of MHD Disturbances around Rotating Black Hole}

\def\rg{\sqrt{-g}}
\def\rI{{r}_{\rm I}}
\def\rF{{r}_{\rm F}}
\def\rH{{r}_{\rm H}}
\def\xF{{x}_{\rm F}}
\def\rhow2{\rho_{\rm w}{}^2}
\def\OmgF{\Omega_{\rm F}}
\def\OmgH{\Omega_{\rm H}}
\def\Omge2{\Omega_{\rm e}{}^2}
\def\SigmaH{\Sigma_{\rm H}}
\def\UrH{U_{\rm H}^r}
\def\UtH{U_t^{\rm H}}
\def\UphiH{U_\phi^{\rm H}}
\def\DeltaF{\Delta_{\rm F}}

\def\Ftph {F_{t      \phi  }}
\def\Frth {F_{r      \theta}}

\def\Mach{{\cal M}}
\def\Ufm{U_{\rm FM}}

\begin{document}

\title{Collimation of Highly Variable Magnetohydrodynamic Disturbances
       around a Rotating Black Hole}

\author{K. Hirotani}  
\affil{National Astronomical Observatory, Osawa 2-21-1, Mitaka,
    Tokyo 181, Japan}


\begin{abstract}
We have studied non-stationary and non-axisymmetric perturbations of 
a magnetohydrodynamic accretion onto a rotating (Kerr) black hole.
Assuming that the magnetic field dominates the plasma accretion, 
we find that the accretion suffers a large radial acceleration 
resulting from the Lorentz force, and becomes highly variable 
compared with the electromagnetic field there. 
In fact, we further find an interesting perturbed structure 
of the plasma velocity with a large peak in some narrow region 
located slightly inside of the fast-magnetosonic surface. 
This is due to the concentrated propagation of the fluid disturbances
in the form of fast-magnetosonic waves along the separatrix surface.
If the fast-magnetosonic speed is smaller in the polar regions
than in the equatorial regions, the critical surface 
has a prolate shape for radial poloidal field lines.
In this case, only the waves that propagate towards the equator
can escape from the super-fast-magnetosonic region 
and collimate polewards as they propagate outwards
in the sub-fast-magnetosonic regions.
We further discuss the capabilities of such collimated waves
in accelerating particles due to cyclotron resonance in
an electron-positron plasma.
\end{abstract}



%

\section{Introduction}

It is commonly accepted that extragalactic jets are intimately linked
with accretion process onto supermassive black holes residing in the 
central regions of active galactic neuclei.
These jets are initially relativistic,
as indicated by superluminal proper motions of radio emitting knots
(e.g., Wehrle et al. 1992) and by high-energy, rapidly variable
$\gamma$-ray emissions (e.g., Montigny et al. 1995).
Moreover, Hobble Space Telescope studies of the base of M87 jet
reveal a rotating gas disk apparently lying normal to the 
jet direction (Ford et al. 1994; Harms et al. 1994).
Despite intense study, the underlying formation mechanism is still
uncertain.
Nevertheless, hydrodynamic and magnetohydrodynamic processes 
associated with the accretion disk seem to be a promising candidate
mechanism.

Possible flows of the energy conversion from the accretion to a
small fraction of gas in jets
have been suggested by Shakura and Sunyaev (1973) 
in the context of a thick supercritical accretion disk which exhibites 
inflow along the equator and outflow near to the poles. 
If the radiation-supported rotating gas adopts a 
hydrostatic toroidal configuration, 
then a pair of funnels are defined which could be responsible for the 
production of jets along the rotational axis (Lynden-Bell 1978). 
Blandford and Payne (1982) considered a magnetized disk
and showed that a gas leaves the disk in a centrifugally driven wind
provided that the magnetic field makes an angle of less than
$60^\circ$ with the radius vector at the disk. 
Furthermore, the magnetohydrodynamic (MHD) disturbances produced at a 
galactic nucleus with a compact nuclear disk would be strongly collimated 
polewards resulting in jets, 
if the Alfv\'en velocity in the disk is much higher than its surroundings 
(Sofue 1980). 
Such a collimation would lead to an increase in the wave amplitude, 
resulting in shock waves which are conjectured to develop into 
a strongly compressed region of the magnetic field. 
In this region, high-energy particles are likely to be accelerated in the 
perpendicular direction to the equatorial nuclear disk. 

The purpose here is to explore the issue
whether MHD waves can convey some portions of accretion energy 
to the polar regions in the vicinity of a black hole.
Causality requires that the MHD inflows should pass through the 
fast-magnetosonic point 
and become super-fast-magnetosonic at the horizon. 
The investigation of the so-called critical condition that the inflow should 
pass through this point smoothly offers, 
in fact, the key to an understanding of MHD interactions in a black hole 
magnetosphere. 

The MHD interactions are expected to work most effectively in the 
magnetically dominated limit in which the rest-mass energy density of 
particles is negligible compared with the magnetic energy density. 
In this limit, the fast-magnetosonic point is located very close to the 
horizon (e.g., Phinney 1983); 
as a result, a general relativistic treatment is required. 
Analyzing the critical condition in a stationary and axisymmetric 
magnetically dominated black hole magnetosphere, 
Hirotani et al. (1992, hereafter Paper I) showed that 
roughly $10\%$ of the rest-mass energy and a significant fraction of the 
initial angular momentum are transported from the fluid to the 
magnetic field during the infall. 
Furthermore, if a small-amplitude perturbation is introduced into the 
magnetosphere, a lot of perturbation energy is deposited
from the magnetic field to the fluid near to the fast-surface
in the short-wavelength limit; 
accordingly, the plasma accretion becomes highly variable there 
(\cite{hir93}, hereafter Paper II). 
Subsequently, Hirotani and Tomimatsu (1994, Paper III)
investigated the spatial structure of the disturbance in a 
Schwarzschild metric, assuming that the characteristic scale 
of the radial variations of perturbed quantities is comparable 
to that of unperturbed quantities instead of adopting the 
short-wavelength limit. 
They revealed that the magnetically dominated accretion becomes
most variable at the fast-magnetosonic separatrix surface 
and that the large-amplitude fluid's disturbance can escape
into the sub-fast-magnetosonic regions by propagating meridionally
almost along the separatrix.
In this paper, we extend the analysis performed in Paper III to a
Kerr metric, further examine the propagation of the escaped fast
waves in the sub-fast regions, and discuss the possibilities of 
particle acceleration due to a collimation of such waves.

The outline of this paper is as follows. 
In \S 2, we formulate non-stationary non-axisymmetric 
perturbations of the MHD accretion and derive the wave equation
describing the perturbation.
Solving the wave equations, we show in \S 3 that 
the fluid becomes most variable slightly inside of the fast-surface,
which is consistent with the results obtained in Paper II and III. 
We further demonstrate that the disturbances can escape 
into the sub-fast regions in the form of fast waves by propagating 
towards the equator provided that the fast-magnetosonic speed, $\Ufm$
is slower in the polar regions than in the equatorial regions.
In \S 4, 
the escaped fast waves will be shown to collimate towards the 
rotational axis under such a distribution of $\Ufm$.
We finally discuss the capabilities of such collimated waves
in accelerating of particles due to nonlinear interactions
between the waves and electron-positron plasmas in \S 5.

\section{Magnetically Dominated Accretion}

We will begin by considering basic equations describing a 
magnetosphere around a rotating black hole. 
Since the self-gravity of the electromagnetic field and plasma 
around a black hole is very weak, 
the background geometry of the magnetosphere is described 
by the Kerr metric, \par

\begin{eqnarray}
  ds^2 &=& {\Delta-a^2\sin^2\theta \over \Sigma} dt^2 
           +{4Mar\sin^2\theta \over \Sigma} dt d\phi
  \nonumber    \\
  &&       -{A \sin^2\theta \over \Sigma} d\phi^2
           -{\Sigma \over \Delta} dr^2 
           -\Sigma d\theta^2,
  \label{eq:metric}
\end{eqnarray}

\noindent
where $\Delta\equiv r^2-2Mr+a^2$, $\Sigma\equiv r^2+a^2\cos^2\theta$, 
$A\equiv (r^2+a^2)^2-\Delta a^2\sin^2\theta$ and $a\equiv J/M$; 
$M$ is the mass of a hole. 
Throughout this paper we use geometrized units such that $c=G=1$. 

Under ideal MHD conditions, since the electric field vanishes 
in the fluid rest frame, we have $ F_{\mu\nu} U^\mu = 0$, 
where $F_{\mu\nu}$ is the electromagnetic field tensor satisfying 
Maxwell equations, $ F_{[\mu\nu,\rho]} = 0 $ and $U^\mu$ is the fluid 
four velocity. 
The motion of the fluid in the cold limit 
is governed by the following equations of motion: 

\begin{eqnarray}
  T^{\mu\nu}{}_{;\nu} &=& 
    \Biggl[ \mu n U^{\mu} U^{\nu} 
          + {1 \over 4\pi} \bigl( F^{\mu\rho}F_{\rho}{}^{\nu}
                                  + {1 \over 4} g^{\mu\nu} 
                                    F_{\alpha\beta}F^{\alpha\beta}
                           \bigr)
   \Biggr]_{;\nu}
  \nonumber \\
  &=& 0,  \label{eq:eq-motion}
\end{eqnarray}

\noindent
where the semicolon denotes a covariant derivative 
and $\mu$ the rest-mass of a particle.
For electron-proton plasmas $\mu$ refers to a rest-mass of a proton,
whereas for electron-positron plasmas $\mu$ refers to that of 
an electron (or a positron).
The proper number density ($n$) obeys the continuity equation, 
$ (nU^{\mu})_{;\mu} = 0 $. 
We adopt these basic equations for a description of 
stationary and axisymmetric black-hole magnetospheres in \S 2.1
and for an analysis of perturbed state in \S 2.2 and afterwards. 

\subsection{Unperturbed Magnetosphere}

From an analysis of the stationary and axisymmetric ideal MHD equations, 
it is known that there exist four integration constants that are 
conserved along each flow line 
(e.g., \cite{bek78}; Camenzind 1986a,b). 
These conserved quantities are the angular velocity of a magnetic field line 
($\Omega_F$), the particle flux per magnetic flux tube ($\eta$), 
the total energy ($E$) and the total angular momentum ($L$) per particle. 
They are defined as follows: 
\par

\begin{equation}
  \Omega_F = {F_{tr} \over F_{r\phi}} = {F_{t\theta} \over 
   F_{\theta\phi}},
  \label{eq:def-OmgF}
\end{equation}

\begin{eqnarray}
  \eta &=& -{\sqrt{-g} n U^r \over F_{\theta\phi}} 
        =  -{\sqrt{-g} n U^{\theta} \over F_{\phi r}} 
  \nonumber \\
  &=& -{\sqrt{-g} n (U^\phi - \Omega_F U^t) \over F_{r\theta}},
  \label{eq:def-eta}
\end{eqnarray}

\begin{equation}
  E \equiv \mu U_t - {\Omega_F \over 4\pi\eta} B_{\phi}
  \label{eq:def-E}
\end{equation}

\noindent
and 

\begin{equation}
  L = -\mu U_{\phi} - {1 \over 4\pi\eta} B_{\phi},
  \label{eq:def-L}
\end{equation}

\noindent
where the toroidal magnetic field ($B_\phi$) is defined by 

\begin{equation}
  B_\phi = - {\rhow2 \over \rg} \Frth;
\end{equation}

\begin{equation}
  \rhow2 \equiv \Delta \sin^2\theta, \qquad
  \rg \equiv \Sigma\sin\theta .
\end{equation}

\noindent
When $0 < \OmgF < \OmgH$ holds, 
thereby hole's rotational energy and angular momentum are extracted
magnetohydrodynamically (Blandford and Znajek 1977; Phinney 1983),
both $E$ and $L$ become negative.
A poloidal flow line is identical with a poloidal magnetic field line, 
and is given by $\Psi(r,\theta)=$ constant, 
where $\Psi$ is the $\phi$-component of the unperturbed 
electromagnetic vector potential. 
The conserved quantities are functions of $\Psi$ alone.

We next describe a stationary plasma accretion in a black-hole magnetosphere. 
In a black-hole magnetosphere there are two light surfaces. 
One is called the outer light surface, 
which is formed by centrifugal force in the same manner as in pulsar models. 
The other is called the inner light surface, which is formed by the 
gravity of the hole. 
In a region between the horizon and the inner light surface, 
the plasma must stream inwards, 
while in a region beyond the outer light surface 
it must stream outwards. 
A plasma source in which both inflows and outflows start with 
a low poloidal velocity is located 
between these two light surfaces (Nitta et al. 1991); 
the injection region ($r=\rI$) of the accretion may be 
a pair-creation zone above the disk 
(Beskin et al. 1992; Hirotani and Okamoto 1997)
or the disk surface whose inner edge corresponds to the 
innermost stable circular orbit in a Kerr geometry. 
Along the magnetic field lines the plasma inflows 
pass through the Alfv\'en point, 
the light surface, and the fast magnetosonic point ($r=\rF$), successively; 
they finally reach the event horizon ($r=\rH$). 
From now on we use the subscripts I, F and H to indicate 
that the quantities are to be evaluated at $r=\rI$, $r=\rF$
and $r=\rH$, respectively. 
In a magnetically dominated magnetosphere,
the fast-magnetosonic point is located very close to the horizon
(for explicit expression of $\rF$ and $\rH$ see Paper I).

\subsection{Perturbed Magnetosphere}

We next consider a small-amplitude non-axisymmetric perturbation 
superposed on the unperturbed state discussed in the last subsection. 
In the perturbed state all perturbation equations are solved 
self-consistently, including the trans-field equation. 
We wish to examine the behavior of fluid quantities 
(energy, angular momentum, and poloidal velocity) 
in response to small variations in the magnetic field at 
various points in the magnetosphere, 
especially at the fast-point.

Let the actual poloidal component of the electric and magnetic fields, 
as a result of the disturbance, be $\,$ 
$E_A+e_A$ and $B_A+b_A$  ($A=r,\theta$), respectively; 
the small letters ($e_r$, $e_\theta$, $b_r$, and $b_\theta$) are the 
Eulerian perturbations of the corresponding quantities. 
Let $e_\phi$ and $b_\phi$ denote perturbations of 
$\Ftph$ and  $-\rg F^r{}_\theta$, respectively. 
Furthermore, we introduce $u^r$ and $u^\theta$ 
such that the actual component of the poloidal fluid velocity field, 
as a result of the disturbance, becomes $ U^A + u^A $. 
Let $u_t$ and $-u_\phi$ be the Eulerian perturbations of 
fluid energy and angular momentum per unit mass, respectively. 

Making use of (4), 
we can simplify the $t$, $\phi$ components of the frozen-in 
conditions as follows: 

\begin{equation}
  U^r e_r + \OmgF \Psi_\theta u^\theta + U^\phi e_\phi = 0,  
  \label{eq:frozen-t}
\end{equation}

\begin{equation}
  -U^t e_\phi + U^r \rg b^\theta - \Psi_\theta u^\theta = 0.
  \label{eq:frozen-phi}
\end{equation}

\noindent
The $\theta$-component of the frozen-in condition in a little bit 
complicated.
To take account of cancelations in the $\Delta$-expansion,
we combine $U^\mu u_\mu = 0$ with the $\theta$-component of the 
frozen-in condition.
We then obtain 

\begin{eqnarray}
  \lefteqn{
    \frac{1}{\rhow2} ( G_1 e_\theta + G_2 \rg b^r ) 
       + g_{rr}g_{\theta\theta} U^r \frac{b_\phi}\rg 
       + \frac{\epsilon}{\mu} \frac{\Psi_\theta}{G_1} u_\phi
          } 
  \nonumber \\ 
  && + \frac{\rg B_\phi [ (\epsilon/\mu) U_\phi 
                          -(g_{t\phi}+g_{\phi\phi}\OmgF) ]}
            {(G_1 + G_2 \OmgF)G_1}
       u^r    = 0. \quad
  \label{eq:frozen-th0}
\end{eqnarray}

\noindent
where

\begin{eqnarray}
G_1 &\equiv&  g_{\phi\phi}U_t    -g_{t\phi}U_\phi,  \\
G_2 &\equiv& -g_{tt}      U_\phi +g_{t\phi}U_t .
\end{eqnarray}

\noindent
Since we are concerned with magnetohydrodynamical interactions 
near to the fast-surface located very close to the horizon,
we evaluate unperturbed quantities appearing in the perturbed
equations in the vicinity of the horizon ($\Delta \ll M^2$).
Then equation (\ref{eq:frozen-th0}) reduces to

\begin{eqnarray}
  \lefteqn{
    \frac{2M\rH \SigmaH \sin^2 \theta}{\Delta} (\UrH)^2
    \left[ \frac{2M\rH}{\SigmaH} h 
           + \frac{b_\phi}{\sin\theta}  \right]
    + \frac{\epsilon}{\mu} \Psi_\theta u_\phi  }
   \nonumber \\
  && - \frac{\Psi_\theta}{\UrH} 
       \left[ \frac{\epsilon}{\mu}U_\phi^{\rm H}
              + (\OmgH-\OmgF)g_{\phi\phi}^{\rm H} 
       \right] u^r = 0,
  \label{eq:frozen-th}
\end{eqnarray}

\noindent
where

\begin{eqnarray}
  \epsilon &\equiv& E - \OmgF L,
  \label{eq:def-eps} \\
  h        &\equiv& e_\theta -\OmgH\rg b^r;
  \label{eq:def-h}
\end{eqnarray}

\noindent
$\OmgH$ is the hole's rotational angular frequency 
and defined by 

\begin{equation}
  \OmgH \equiv -g_{t\phi}^{\rm H}/g_{\phi\phi}^{\rm H}
         =  -g_{tt}^{\rm H}/g_{t\phi}^{\rm H} .
\end{equation}

\noindent
Here use has been made of the fact that unperturbed fluid velocity 
$\UrH$ is related with fluid energy $\UtH$ and angular momentum
$-\UphiH$ as (Paper I)

\begin{equation}
  \UrH = - \frac{2M\rH}{\SigmaH} ( U_t^{\rm H} + \OmgH U_\phi^{\rm H} )
         + O(\Delta/M^2).
  \label{eq:UrH}
\end{equation}

\noindent
$\UrH$ and hence $\UtH$ and $-\UphiH$, are of order of unity.
The explicit expressions of these quantities are given in Paper I.

If we assume that the characteristic scales of {\it meridional} variations 
in the perturbed state are much shorter than those in the 
unperturbed state, 
the perturbation equations reduce significantly. 
For non-axisymmetric perturbations, all perturbed quantities may
therefore be assumed to be proportional to 
$\exp(i\omega t -ik_\theta \theta -im\phi)$, 
where $\vert k_\theta \vert \gg 1$. 
Under this approximation, we obtain from the homogeneous part of 
Maxwell equations

\begin{equation}
  im b_\phi 
  + \Delta \sin^2\theta 
    \left( -i k_\theta b^\theta 
           +\frac{2M\rH}\Delta \frac{db^r}{dr_*}  \right)
 = 0,
  \label{eq:max-t}     
\end{equation}

\begin{equation}
  \frac{2M\rH}\Delta \frac{de_\phi}{dr_*}
  + im e_r + i\omega\rg b^\theta = 0, 
  \label{eq:max-th}
\end{equation}

\begin{equation}
  ik_\theta e_r +\frac{2M\rH}\Delta \frac{de_\theta}{dr_*}
  + i\omega \frac{\Sigma_{\rm H}}{\Delta\sin\theta} b_\phi = 0,
  \label{eq:max-phi}
\end{equation}

\noindent
where $r_*$ is the tortoise coordinate defined by 

\begin{equation}
  {dr_* \over dr} \equiv \frac{r^2+a^2}{\Delta}.
  \label{eq:def-tort}
\end{equation}

\noindent
It is convenient to introduce this coordinate when we describe waves near to 
the horizon, because the interval $(r_{\rm H},\infty)$ in the $r$-coordinate 
is stretched to $(-\infty,\infty)$ in $r_*$. 
We assume that the characteristic scale of 
the radial variations is comparable to that of the gravitational field, 
that is, $  \vert df/dr_* \vert \approx \vert f/M \vert $, 
where $f$ denotes some perturbed quantity. 
In this paper we do not take the short-wavelength limit 
($ \vert df/dr_* \vert \gg \vert f/M \vert $) which was adopted in Paper II.

Eliminating $u^\theta$ from equations (\ref{eq:frozen-t}) and 
(\ref{eq:frozen-phi}),
we obtain a relation between $e_r$, $b^\theta$, and $e_\phi$.
Combining this relation with equation (\ref{eq:max-th}),
we can express $e_r$ and $b^\theta$ in terms of $e_\phi$ as follows:

\begin{equation}
  (\omega-m\OmgF)e_r = -i\OmgF \frac{2M\rH}{\Delta}
     \left[ \frac{de_\phi}{dr_*} 
            + i\omega \frac{\OmgH-\OmgF}{\OmgF} e_\phi  \right],
  \label{eq:elemag-3}
\end{equation}

\begin{eqnarray}
  \lefteqn{ (\omega-m\OmgF)\rg b^\theta }  \nonumber \\
  && = i \frac{2M\rH}{\Delta}
     \left[ \frac{de_\phi}{dr_*} + im (\OmgH-\OmgF) e_\phi  \right].
  \label{eq:elemag-4}
\end{eqnarray}

\noindent
Substituting equation (\ref{eq:elemag-3}) into equation 
(\ref{eq:max-phi}), we have

\begin{eqnarray}
  \frac{de_\theta}{dr_*} 
  &=& - \frac{k_\theta}{\omega-m\OmgF}
        \left[ \OmgF\frac{de_\phi}{dr_*} 
               + i\omega(\OmgH-\OmgF)e_\phi \right] 
  \nonumber \\
  && -i\omega \frac{\SigmaH}{2M\rH} \frac{b_\phi}{\sin\theta}
  \label{eq:elemag-5}
\end{eqnarray}

\noindent
In the same manner, equations (\ref{eq:max-t}) and (\ref{eq:elemag-4})
give

\begin{eqnarray}
  \rg\frac{db^r}{dr_*} 
  &=&  - \frac{k_\theta}{\omega-m\OmgF}
         \left[ \frac{de_\phi}{dr_*} + im(\OmgH-\OmgF)e_\phi \right] 
  \nonumber  \\
  && -im \frac{\SigmaH}{2M\rH} \frac{b_\phi}{\sin\theta}
  \label{eq:elemag-6}
\end{eqnarray}

\noindent
Equations (\ref{eq:elemag-3})-(\ref{eq:elemag-6}) express 
poloidal components of perturbed electromagnetic fields 
in terms of their toroidal components.

Let us next consider the equations of motion. 
First, the definition of proper time $U^\mu u_\mu =0$ yields near the 
horizon with the aid of equation (\ref{eq:UrH}), 

\begin{equation}
  u^r = -\frac{2M\rH}{\Sigma_H}(u_t+\OmgH v_\phi) + O(\Delta/M^2).
  \label{eq:fluid-1}
\end{equation}

\noindent
Secondly, the $t$-component of the equation of motion gives

\begin{equation}
  (\mu n U^\mu \nabla_\mu U_t)^{(1)} + f_t^{(1)} = 0 ; 
  \label{eq:EOM-t-0}
\end{equation}

\noindent
in the vicinity of the horizon we have

\begin{equation}
  (U^\mu \nabla_\mu U_t)^{(1)} = \frac{2M\rH}{\Delta}\UrH\Lambda u_t,
\end{equation}

\begin{equation}
  \Lambda = -i(\omega-m\OmgH) + \frac{d}{dr_*} ,
  \label{eq:def-Lambda}
\end{equation}

\begin{eqnarray}
  \lefteqn{ f_t^{(1)} = \frac{\OmgF \Psi_\theta}{4\pi\SigmaH}
                        \frac{2M\rH}{\Delta} }  \nonumber \\
  && \left[ i(\omega-m\OmgH)\frac{2M\rH}{\SigmaH}h
            + \frac{d}{dr_*} \left( \frac{b_\phi}{\sin\theta} \right)
     \right] .
  \label{eq:def-ft}
\end{eqnarray}

\noindent
The subscript $(1)$ indicates that the quantity is evaluated in the 
perturbed state.
In general, the $\alpha$-component of the Lorentz force is defined by

\begin{equation}
  f_\alpha \equiv \frac{ F_\alpha{}_\mu \partial_\nu
                         (\rg F^{\mu\nu})}
                       {4\pi\rg} .
  \label{eq:def-falpha}
\end{equation}

\noindent
Setting $\alpha=t$, taking the linear order in the perturbation,
and evaluating in the vicinity of the horizon,
we obtain equation (\ref{eq:def-ft}).

The perturbed fluid's density, $n_1$, can be eliminated
from equation (\ref{eq:EOM-t-0}) with the aid of the continuity
equation

\begin{equation}
  \UrH \Lambda \left( \frac{n_1}{n_0} \right) + H(u) = 0,
  \label{eq:cont-eq}
\end{equation}

\noindent
where

\begin{eqnarray}
  H(u) &\equiv& 
          \frac{du^r}{dr_*}
          + i(\omega-m\OmgH) \frac{2M\rH}{\SigmaH}(u_t+\OmgH u_\phi)
  \nonumber \\
  &&  - ik_\theta \frac{\Delta}{2M\rH} v^\theta .
  \label{eq:def-H}
\end{eqnarray}

\noindent
Operating $\Lambda$ to the both sides of equation (\ref{eq:cont-eq}),
and combining with (\ref{eq:EOM-t-0}) to eliminate $n_1$, we obtain

\begin{eqnarray}
  && \hspace{-0.5 truecm}
      \mu n_0 \left\{ \frac{2M\rH}{\Delta}\UrH  
                     \left( -\frac{\rH-M}{M\rH} +\Lambda \right) u_t
                     - (\partial_r U_t)H(u) \right\} 
  \nonumber \\
  && \qquad + \Lambda f_t^{(1)} = 0
  \label{eq:EOM-t}
\end{eqnarray}

\noindent
In the same manner, from the $\phi$-component of the equation of motion,
we obtain 

\begin{eqnarray}
  && \hspace{-0.5 truecm}
       \mu n_0 \left\{ \frac{2M\rH}{\Delta}\UrH  
                    \left( -\frac{\rH-M}{M\rH} +\Lambda \right) u_\phi
                    - (\partial_r U_\phi)H(u) \right\} 
  \nonumber \\
  && + \Lambda f_\phi^{(1)} = 0,
  \label{eq:EOM-phi}
\end{eqnarray}

\noindent
where

\begin{equation}
  f_\phi^{(1)} = -\frac{1}{\OmgF}f_t^{(1)} +O(\Delta/M^2) .
\end{equation}

Finally, $\theta$-component of the equation of motion becomes

\begin{equation}
  (\mu n U^\mu \nabla_\mu U_\theta)^{(1)} + f_\theta^{(1)} = 0, 
  \label{eq:EOM-th-0}
\end{equation}

\noindent
where 

\begin{eqnarray}
  \lefteqn{ f_\theta^{(1)} = -\frac{2M\rH}{4\pi\Delta\sin\theta} 
            \left\{ \left( \frac{2M\rH}{\SigmaH} \right)^2 
                    (\OmgH-\OmgF)\Psi_\theta    \right. } 
  \nonumber \\
  && \times \left[ \Lambda(e_r+\OmgH\rg b^\theta)
                  -\frac{ik_\theta}{2M\rH}h \right]    
  \nonumber \\
  && \left. -\frac{ik_\theta}{2M\rH}B_\phi \frac{b_\phi}{\sin\theta}
            -i\frac{\omega-m\OmgF}{2M\rH} \frac{\Psi_\theta}{\sin\theta}
             e_\phi   \right\}
  \label{eq:def-f1th}
\end{eqnarray}

\noindent
Eliminating $n_1$ with the aid of equation (\ref{eq:cont-eq}),
taking the leading orders in $\Delta$-expansion and 
considering relative amplitude between perturbed quantities
together with the dispersion relation for the outgoing 
fast-magnetosonic mode (Paper II),
we can self-consistently neglect fluid contributions in equation
(\ref{eq:EOM-th-0}) to obtain

\begin{equation}
  \Lambda f_{(1)}^\theta = 0.
  \label{eq:EOM-th-1}
\end{equation} 

\noindent
Using equations (\ref{eq:frozen-th}), (\ref{eq:elemag-3}), 
(\ref{eq:elemag-4}),
we can rewrite equation (\ref{eq:EOM-th-1}) into

\begin{eqnarray}
  \lefteqn{ \frac{(2M\rH)^2}{\Delta\SigmaH}
            \frac{\OmgH-\OmgF}{\omega-m\OmgF}
            \left( -\frac{\rH-M}{M\rH} + \Lambda \right)
            \Lambda e_\phi  }  \nonumber \\
  && -\frac{k_\theta}{2M\rH} 
      \left[ \frac{2M\rH}{\SigmaH}h + \frac{b_\phi}{\sin\theta} \right]
     = 0.
  \label{eq:EOM-th1}
\end{eqnarray}

We analyze a system which is formed by $10$ equations 
(equations [\ref{eq:frozen-phi}], [\ref{eq:frozen-th}],
           [\ref{eq:elemag-3}]-[\ref{eq:fluid-1}],
           [\ref{eq:EOM-t}],[\ref{eq:EOM-phi}], and
           [\ref{eq:EOM-th1}])
in $10$ unknown functions 
($b^r$, $b^\theta$, $e_r$, $e_\theta$, $b_\phi$, $e_\phi$, 
$u_t$, $u^r$, $u^\theta$, and $u_\phi$). 
The perturbed fluid density would be calculated 
if we solved the perturbed continuity equation. 
These 10 equations give some relations between 
two arbitrary perturbed quantities, 
and are further combined into a single differential equation. 

Substituting equation (\ref{eq:elemag-4}) into equation 
(\ref{eq:frozen-phi}), we obtain

\begin{equation}
  (\omega-m\OmgF)\Psi_\theta u^\theta 
  = i \frac{2M\rH}{\Delta} \UrH \Lambda e_\phi
  \label{eq:EQ1}
\end{equation}

\noindent
We can use this equation to eliminate $u^\theta$ and examine
the relative amplitude between $u^\theta$ and $e_\phi$. 

One combination of equations (\ref{eq:EOM-t}) and (\ref{eq:EOM-phi})
yields, with the aid of equation (\ref{eq:fluid-1}),

\begin{eqnarray}
  && \OmgF\Psi_\theta \frac{\mu}{E} \Lambda u^r
            = -i(\omega-m\OmgH) \frac{2M\rH}{\SigmaH}h 
  \nonumber \\
  && -\frac{d}{dr_*} \left( \frac{b_\phi}{\sin\theta} \right)
  \label{eq:EQ6}
\end{eqnarray}

\noindent
Here, use has been made of the fact that the unperturbed fluid quantity

\begin{equation}
  U_t +\OmgF U_\phi = \frac{\epsilon}{\mu} \equiv \frac{E-\OmgF L}{\mu}
\end{equation}

\noindent
is constant along each flow line. 
The other combination of equations (\ref{eq:EOM-t}) and
(\ref{eq:EOM-phi}) yields

\begin{equation}
  u_t = -\OmgF u_\phi + O(\Delta/M^2)
 \label{eq:ut-uphi}
\end{equation}

Using equation (\ref{eq:fluid-1}) and (\ref{eq:ut-uphi}),
we can eliminate $u_\phi$ in equation (\ref{eq:frozen-th}) to obtain

\begin{eqnarray}
  && -\frac{K_{\rm I}-K_{\rm H}}{\OmgH-\OmgF}\Psi_\theta u^r
            = \frac{\SigmaH}{\Delta}(\UrH)^3 2M\rH \sin^2\theta
  \nonumber \\
  && \qquad \times 
     \left[ \frac{2M\rH}{\SigmaH} h + \frac{b_\phi}{\sin\theta} 
     \right],
  \label{eq:EQ10}
\end{eqnarray}

\noindent
where the effective potential $K$ is defined
by (Takahashi et al. 1990)

\begin{equation}
  K \equiv g_{tt} +2g_{t\phi}\OmgF +g_{\phi\phi}\OmgF^2 .
  \label{eq:def-K}
\end{equation}

\noindent
At the injection point where $U^r$ vanishes,
$K$ takes a positive value, $K_{\rm I}=(\epsilon/\mu)^2$,
whereas at the horizon we have 
$K=K_{\rm H}=g_{\phi\phi}^{\rm H} (\OmgH-\OmgH)^2 < 0$.
Moreover, equations (\ref{eq:elemag-5}) and (\ref{eq:elemag-6}) give

\begin{eqnarray}
  \lefteqn{ -\frac{dh}{dr_*}
            = \frac{(\OmgH-\OmgF)k_\theta}{\omega-m\OmgF} \Lambda e_\phi }
  \nonumber \\
  && -i(\omega-m\OmgH) \frac{\SigmaH}{2M\rH} \frac{b_\phi}{\sin\theta}
  \label{eq:EQ11}
\end{eqnarray}

So far, we have obtained four independent equations
(\ref{eq:EOM-th1}), (\ref{eq:EQ6}), (\ref{eq:EQ10}), and 
(\ref{eq:EQ11}) for four unknowns $u^r$, $h=e_\theta-\OmgH\rg b^r$,
$e_\phi$, and $b_\phi$.
Combining these four equations,
we finally obtain the wave equation

\begin{eqnarray}
  \lefteqn{ 
     \left\{ (\Delta-\DeltaF)\frac{d^2}{dr_*{}^2} \right. }
  \nonumber \\
  && + \left[ \frac{\rH-M}{M\rH}(\Delta+\DeltaF)
              +2i(\omega-m\OmgH)\DeltaF \right] \frac{d}{dr_*} 
  \nonumber \\
  && -i(\omega-m\OmgH) \left[ \frac{\rH-M}{M\rH}+i(\omega-m\OmgH) \right]
       (\Delta+\DeltaF)
  \nonumber \\
  && - \left. \left( \frac{\Delta}{2M\rH} \right)^2 k_\theta{}^2 
       \right\}  u^r  = 0,
  \label{eq:master-1}
\end{eqnarray}

\noindent
where $\DeltaF$ is given by (Paper I)

\begin{equation}
  \DeltaF = \frac{2M\rH\SigmaH\sin^2\theta\OmgF(\OmgH-\OmgF)(\UrH)^3}
                 {K_{\rm I}-K_{\rm H}}
            \frac{\mu}{E}.
  \label{eq:def-deltaF}
\end{equation}

\noindent
Introducing a new non-dimensional radial coordinate as

\begin{equation}
  x \equiv \frac{\Delta}{\Sigma},
  \label{eq:def-x}
\end{equation}

\noindent
and recovering meridional derivatives by setting
$k_\theta = i\partial_\theta$,
we can rewrite equation (\ref{eq:master-1}) into

\begin{eqnarray}
  \lefteqn{ x^2(x-\xF) \frac{\partial^2 u^r}{\partial x^2}
            + 2x(x+i\sigma\xF) \frac{\partial u^r}{\partial x} }
  \nonumber \\
  && + \left[ -i\sigma(1+i\sigma)(x+\xF)
              +\frac{\SigmaH \, x^2}{4(\rH-M)^2} 
               \frac{\partial^2}{\partial\theta^2} \right] 
       u^r   \nonumber \\
  && = 0  
  \label{eq:master-2}
\end{eqnarray}
  
\noindent
where non-dimensional corotational frequency $\sigma$ is defined by

\begin{equation}
  \sigma \equiv (\omega-m\OmgH)\frac{M\rH}{\rH-M}.
  \label{eq:def-sigma}
\end{equation}

\noindent
Equation (\ref{eq:master-2}) becomes elliptic in the sub-fast region
($x>\xF$), while it becomes hyperbolic in the super-fast region
($x<\xF$).

From equations (\ref{eq:fluid-1}) and (\ref{eq:ut-uphi}),
we can see that fluid's energy ($u_t$) and angular momentum ($-u_\phi$)
obey the same equation as (\ref{eq:master-2}).
Therefore, equation (\ref{eq:master-2}) describes the fluid's disturbance
near to the horizon.
In the slowly rotating limit ($a \rightarrow 0$),
equations (\ref{eq:def-x})-(\ref{eq:def-sigma}) reduces to equation
(25) in Paper III,
in which axisymmetric ($m=0$) perturbations in a Schwarzshild metric 
were examined.

\section{Highly Variable Accretion} 

To examine the spatial structure of $u^r(x,\theta)$, 
let us rewrite (\ref{eq:master-2}) as

\begin{equation}
  \left\{ \left( -1 -i\sigma +x\frac{\partial}{\partial x}
                    \right) D_{\rm FM}  
                    + \frac{\SigmaH \, x^2}{4(\rH-M)^2}
                      \frac{\partial^2}{\partial\theta^2}
            \right\} = 0
  \label{eq:master-3}
\end{equation}

\noindent
where $D_{\rm FM}$ refers to the differential operator associated with
outgoing-fast-magnetosonic mode and is defined by

\begin{equation}
  D_{\rm FM} \equiv 
     x(x-\xF) \frac{\partial}{\partial x} +i\sigma(x+\xF)+x
  \label{eq:def-Dfm}
\end{equation}

\noindent
Let us examine the radial($x$) dependence of $u^r$ 
by neglecting the $\theta$-derivative term as a first step. 
Under this assumption, the ingoing and the outgoing modes in 
equation (\ref{eq:master-3}) can be completely separated.
Equation $D_{\rm FM}u^r=0$ gives a solution corresponding to the
outgoing mode,

\begin{equation}
  u^r = C_1 \frac{x^{i\sigma}}{[x-\xF(\theta)]^{1+2i\sigma}},
  \label{eq:sol-0}
\end{equation}

\noindent
where $C_1$ is an integration constant. 
In order that the right-hand-side may not diverge at $x=\xF(\theta)$, 
the real part of $1+2i\sigma$ should be non-positive; 
this indicates that outgoing radial waves must decay, 
because no steady supply of perturbation energy across the fast-surface 
$x=\xF(\theta)$ is possible. 
In this paper, we postulate a steady excitation of perturbation 
induced by plasma injection from the equatorial disk 
or a pair production zone above the disk. 
It requires that $\sigma$ should be real. 
Then, to avoid divergence at $x=\xF$, 
we must consider the $\theta$-derivative term in (\ref{eq:master-3})
at least near to the fast-surface. 

Let us modify the solution (\ref{eq:sol-0}) into the form

\begin{equation}
  u^r = C_1 \frac{x^{i\sigma}}
                 {[x-\xF(\theta)+\delta(\theta)]^{1+2i\sigma}}, 
  \label{eq:approx-sol}
\end{equation}

\noindent
where $\delta$ should be a complex function of $\theta$, 
so that we may obtain a regular solution $u^r$ for real $\sigma$. 
Inserting (\ref{eq:approx-sol}) into (\ref{eq:master-3})
and evaluating the equation in the limit 
$\mid (x-\xF+\epsilon)/ \xF \mid \ll 1$, 
we obtain a non-linear first-order differential equation for $\delta$, 

\begin{equation}
  \delta = \left( \frac{d \xF}{d\theta} - \frac{d\delta}{d\theta} 
           \right)^2 .
  \label{eq:eq-delta}
\end{equation}

\noindent
As boundary conditions, we impose the following symmetry conditions,

\begin{eqnarray}
  && \frac{d {\rm Re}(\delta)}{d\theta} 
   = \frac{d {\rm Im}(\delta)}{d\theta}= 0
  \qquad {\rm at} \quad \theta = 0,
  \label{eq:BD00} \\
  && \frac{d {\rm Re}(\delta)}{d\theta} 
   = \frac{d {\rm Im}(\delta)}{d\theta}= 0
  \qquad {\rm at} \quad \theta = \frac{\pi}{2}.
  \label{eq:BD90}
\end{eqnarray}

\noindent
As an example, numerical solutions satisfying these conditions
for $\xF= 0.1 \cos^2\theta + 0.01$ 
(i.e., prolate shape of the fast-surface) are depicted in
Figs.\ \ref{figA} and \ref{figB}.
In Fig. \ref{figA}, 
${\rm Re}(\delta)=(d\xF/d\theta)^2=0.01\sin^2(2\theta)$,
which approximately satisfy equation (\ref{eq:eq-delta})
when $d^2\xF/d\theta^2 \approx 0$ 
(i.e., at $\theta \approx \pi/4$),
is depicted by the dashed line for comparison.
It should be noted that the maximum amplitude surfaces,
$x=\xF-{\rm Re}(\delta)$,
where $u^r$ has a sharp peak,
is located slightly within the fast surface, $x=\xF$.

\def\FigA{       
\begin{figure} 
\centerline{ \epsfxsize=8cm \epsfbox[50 200 500 600]{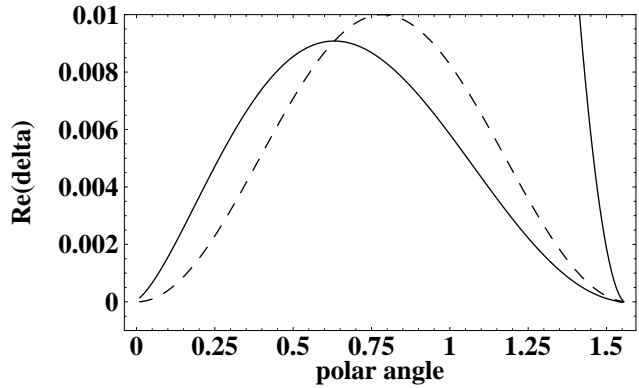} } 
\caption{\label{figA} 
Variation of the real part of $\delta$ as a function of $\theta$. 
The dashed lines, which plots $0.01\sin^2 2\theta$,
is depicted for comparison (see text).
        }
\end{figure} 
}

\FigA

What is most important here is that this maximum amplitude surface
coincides with the separatrix of characteristics.
To see this, it is convenient to introduce a new radial 
coordinate $\xi$ which denotes a deviation from the 
maximum-amplitude surface,

\begin{equation}
  \xi \equiv x-\xF +{\rm Re}(\delta)
  \label{eq:def-xi}
\end{equation}

\noindent
The characteristics of equation (\ref{eq:master-3}) are expressed as

\begin{equation}
  \frac{dx}{d\theta} = \mp \sqrt{\xF(\theta)-x}
  \label{eq:charac-eq}
\end{equation}

\noindent
In the super-fast region, any waves must propagate inwards, $dx<0$.
Therefore, waves propagating to lower latitudes ($d\theta>0$) 
are indicated by the upper sign, 
while those to higher latitudes are indicated by the lower sign.
Combining (\ref{eq:eq-delta}), (\ref{eq:def-xi}), 
and (\ref{eq:charac-eq}), we obtain an equation expressing 
how a characteristic deviates from the maximum-amplitude surface,

\begin{equation}
  \frac{d\xi}{d\theta} 
  =  \mp \sqrt{{\rm Re}(\delta)-\xi}
     - \frac{d\xF/d\theta}{\vert d\xF/d\theta \vert} 
       \sqrt{{\rm Re}(\delta)}.
  \label{separat}
\end{equation}

\noindent
It follows that for a {\it prolate} shape of the fast-surface 
($d\xF/d\theta < 0$), 
$d\xi$ has the same sign as $\xi$ for waves propagating
to lower latitudes ($d\theta > 0$, upper sign).
In other words, outside of the maximum amplitude surface,
characteristics deviate from this surface outwards,
whereas inside of this surface, they deviate inwards.
Poleward propagating waves ($d\theta < 0$, lower sign), 
on the other hand, always
deviate inwards ($d\xi < 0$) to be swallowed by the hole.
That is, only waves propagating towards the equator can escape
into to sub-fast regions, 
if the fast-surface is prolate (Fig.\ \ref{figSide}).
Thus we can regard the maximum amplitude surface as
separatrix of characteristics.
The same discussion could be applied for a oblate shape of the 
fast-surface.

\def\FigB{       
\begin{figure} 
\centerline{ \epsfxsize=8cm \epsfbox[50 200 500 600]{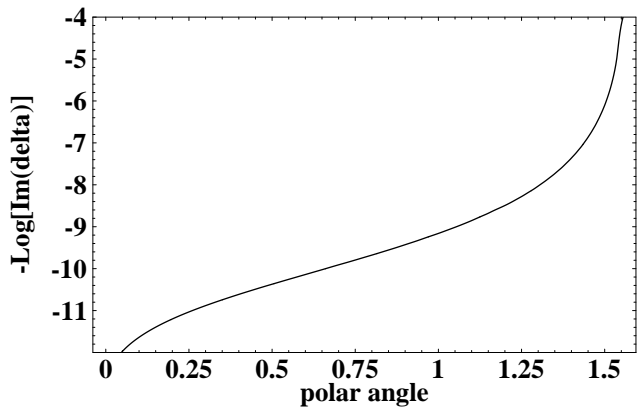} } 
\caption{\label{figB} 
  Logarithmic variation of $1/{\rm Im}(\delta)$
  as a function of $\theta$. 
        }
\end{figure} 
}

\FigB

\def\FigSide{
\begin{figure} 
\centerline{ \epsfxsize=8cm \epsfbox[20 50 350 430]{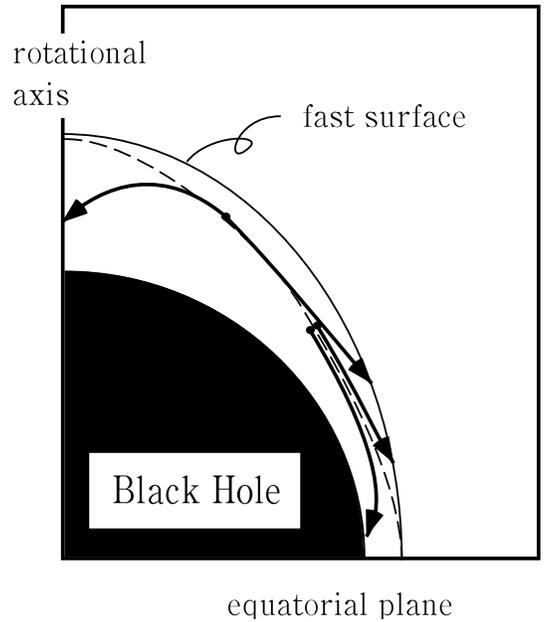} } 
\caption{\label{figSide}
Schematic figure (side view) of a black hole magnetosphere. 
The thick solid curves denote the characteristics of equation 
(62). 
The fast-magnetosonic separatrix surface is denoted by the dashed line. 
The magnetic field lines are not drawn in order to avoid complication. 
The super-fast region is very thin 
in the magnetically dominated limit; 
however, the width of this region is exaggerated. 
        }
\end{figure} 
}

\FigSide

Since fluid obtains most of its perturbation energy from the 
electromagnetic field at the maximum amplitude surface, $\xi=0$,
most of the fluid's disturbance propagates almost along the
separatrix (i.e., the maximum amplitude surface)
and at last deviates inwards or outwards.
In other words, meridional propagation is essential to examine
the spatial structure of fluid disturbance near to the fast-surface.
As a result of the deviation of waves, 
$1/{\rm Im}(\delta)$, and hence fluid amplitude, rapidly
decreases with $\theta$ as indicated in Fig.\ \ref{figB}
for a prolate shape of the fast-surface.
We can quantitatively understand this behavior by considering an 
approximate solution

\begin{equation}
  \rm{Im}(\delta) \propto 
     \exp \left( -\frac12 \int_0^\theta \left[ \frac{d\xF}{d\theta}
                                        \right]
                          d\theta
          \right) ,
  \label{eq:imaginary}
\end{equation}

\noindent
which is applicable when $d^2\xF/d\theta^2 \approx 0$. 
Equation (\ref{eq:imaginary}) indicates that 
$\rm{Im}(\delta)$ decreases exponentially as $\theta$ decreases.

Let us finally consider the relation of amplitude among various 
perturbed quantities.
Combining equations presented in the last section, 
we obtain near to the fast-point 

\begin{equation}
  \frac{e_r}{E_\theta / M}, \frac{b_\theta}{B^r}, u^\theta
  \approx \sqrt{\frac{\mu}{E}} u^r \ll u^r 
  \label{eq:rel-amp}
\end{equation}

\noindent
in order of magnitude because of 
$k_\perp \approx \sqrt{E/\mu} k_{r*}$.
Other electromagnetic quantities have much smaller amplitude
near to the fast-surface.
It is interesting to note that the equipartition of energy 
($[e_r]^2 \approx \mu n [u^r]^2$) is achieved 
near to the fast-surface, 
although the perturbation energy is supplied mainly in the form of 
electromagnetic disturbances 
($[e_r]^2 \approx [E/\mu]\mu n[u^r]^2 \gg \mu n [u^r]^2$) 
far from the horizon, 
In other words, a lot of perturbation energy is deposited from the 
electromagnetic field to the fluid during the infall 
as a result of effective MHD interactions. 
In a realistic magnetically dominated black hole magnetosphere, 
the fluid becomes highly variable ($[u^r]^2 \approx [U^r]^2$) 
near to the fast-surface for a very small input of perturbation energy 
($[e_r]^2 \approx [\mu/E] [E_\theta/M]^2$) from the surroundings. 
The existence of the horizon is essential to make the fluid be
highly variable near to the fast-surface (PaperI, and II).
If the fast-surface is located far from the hole, 
fluid quantities does not become highly variable.

So far, we have derived the following conclusions:
(1) The fluid quantities become highly variable
near to the fast-surface owing to MHD interactions near the horizon.
(2) Their amplitude is a peaking function at the separatrix surface
located slightly inside of the fast-surface.
This is because a large-amplitude fluid disturbance propagates
along the separatrix as a outgoing fast-wave.
(3) For prolate shape of the fast-surface, for instance,
The fast-waves that propagate towards the equator can reach
the fast-surface and escape to the sub-fast region.
These results drive us to the question how these fast-waves
propagate in the sub-fast regions.
In the next section, we will be concerned with this issue.

\section{Wave propagation in sub-fast region}

\subsection{Geometrical optics in relativistic accretion}

As we have seen, meridional wavelength is much less than the radius of
curvature near to the fast surface.
It follows that the propagation of wave packets of fast-magnetosonic
mode follows the laws of geometric optics.
In geometrical optics, the nature of the second-order partial differential
equations describing the propagation can be well studied 
by the characteristic hypersurfaces of the system.
The characteristic hypersurfaces, which play the role of wave fronts,
can be expressed by a surface of
$\psi(x^\mu)=$ constant, where $\psi$ satisfies
the following eikonal equation in the cold limit
(Lichnerowicz 1967; see also Takahashi et al. 1990 for sound waves):

\begin{equation}
  H \equiv s^{\alpha\beta} \psi_{,\alpha} \psi_{,\beta} = 0,
  \label{eq:eikonal-eq}
\end{equation}

\noindent
where $s^{\alpha\beta}$ is defined by

\begin{equation}
  s^{\alpha\beta} \equiv g^{\alpha\beta}
                          + \frac{{U^\alpha}{U^\beta}}{\Ufm^2};
  \label{eq:def-sound-metric}
\end{equation}

\noindent
$\Ufm$ is the fast-magnetosonic speed in $d\tau$-basis and is 
defined by

\begin{eqnarray}
  \Ufm^2 &\equiv& \frac{K B_{\rm p}^2 + B_\phi^2 / \rhow2}
                     {4\pi\mu n}
  \nonumber \\
  &=& \frac{K_{\rm I} - K}{4\pi\mu\eta}
     \frac{B^r}{U^r}.
  \label{eq:def-Ufm}
\end{eqnarray}

\noindent
The first term in equation (\ref{eq:def-sound-metric})
describes the influence of the gravitational field, 
while the second term that of cold, relativistic magnetohydrodynamic
flows.
If we were to replace $\psi_{,\alpha}$ with $k_\alpha$,
we would obtain the dispersion relation for the 
fast-magnetosonic mode,

\begin{equation}
  \left( U^\mu k_\mu \right)^2 + \Ufm^2 k^\mu k_\mu = 0. 
  \label{eq:disp-rel}
\end{equation}

Instead of solving the partial differential equation 
(\ref{eq:eikonal-eq}),
we can investigate the trajectories of wave packets
by solving a set of the following ordinary differential equations:

\begin{eqnarray}
  \frac{dx^\alpha}{d\lambda} 
  &=&  \frac{\partial H(x^\beta,p_\beta)}{\partial p_\alpha},
  \label{eq:HJeq-1} \\
  \frac{dp_\alpha}{d\lambda} 
  &=& -\frac{\partial H(x^\beta,p_\beta)}{\partial x^\alpha},
  \label{eq:HJeq-2}
\end{eqnarray}

\noindent
where $\lambda$ is the parameter along a ray path.
Since the Hamiltonian $H$ contains neither $t$ nor $\phi$,
both wave frequency $\omega=P_t$ and azimuthal wave number
$m= -P_\phi$ are conserved along a ray path.

The unperturbed fluid's velocity field,
on which the wave packets propagate,
must be solved consistently with 
the equations of motion.
First, the definition of proper time gives the poloidal wind 
equation

\begin{equation}
  g_{rr} (U^r)^2 + 1
  = -\frac{g_{\phi\phi}(U_t)^2 -2g_{t\phi}U_tU_\phi +g_{tt}(U_t)^2}
          {\rhow2}
  \label{eq:pol-eq}
\end{equation}

\noindent
Secondly, combining the unperturbed continuity equation,
Maxwell equations, and the frozen-in conditions with equations
(\ref{eq:def-E}) and (\ref{eq:def-L}), we obtain
(Camenzind 1986b)

\begin{eqnarray}
  \mu U_t &=& \frac{ (g_{tt}+g_{t\phi}\OmgF)\epsilon -\Mach^2 E}
                   { K - \Mach^2 },
  \label{eq:exp-Ut} \\
  \mu U_\phi &=& 
         \frac{ (g_{t\phi}+g_{\phi\phi}\OmgF)\epsilon +\Mach^2 L}
                   { K - \Mach^2 },
  \label{eq:exp-Uphi}
\end{eqnarray}

\noindent
where the Alfv$\acute{\rm e}$nic Mach number $\Mach$ is defined as

\begin{equation}
  \Mach^2 \equiv \frac{4\pi\mu\eta^2}{n}
          =      \frac{4\pi\mu\eta}{B^r} U^r.
  \label{eq:def-Mach}
\end{equation}

\noindent
We assume an appropriate functional form for $B^r$ 
instead of solving the unperturbed trans-field equation.

These three equations, (\ref{eq:pol-eq})-(\ref{eq:exp-Uphi}),
together with (\ref{eq:def-Mach}),
give the fluid's velocity field ($U_t$, $U^r$, $-U_\phi$)
on the poloidal plane.
We assume that the accretion along each radial field line
starts from the point at which $K$ becomes 0.55.
This condition defines a nearly spherical (but somewhat oblate)
injection surface of accretion at $\rI \approx 5M$ 
for a mildly rotating hole ($a \approx 0.5M$).
We suppose that there is no flow of plasmas outside of the 
injection surface.
This assumption alters the propagation of the fast waves 
negligibly, because plasma flows in the region 
$5M<r<10M$ is non-relativistic, whereas the fast-magnetosonic speed
in $dt$-basis is slightly smaller than that of light.
We further assume that the energy density of the magnetic field 
is nine times larger than that of the fluid's rest-mass
in the sense that $E/(\mu\sin^2\theta)= -10$.

In this section, 
we trace the ray paths of the fast-magnetosonic wave packets
radiated meridionally with momentum 
$\vert k_\perp \vert \approx \sqrt{E/\mu} \, \vert k_{r*} \vert$
from the fast-magnetosonic surface
rather than radiated spherically (i.e., in all directions),
by solving equations (\ref{eq:HJeq-1}) and (\ref{eq:HJeq-2})
in the magnetically dominated accretion described 
by equations (\ref{eq:pol-eq})-(\ref{eq:exp-Uphi}).

\subsection{Collimation of MHD waves}

Let us first demonstrate typical results when the fast-magnetosonic
speed, $\Ufm$ is slower in polar regions than in equatorial regions.
Such a distribution of $\Ufm$ will be realized when $\vert B^r \vert$
is smaller in the polar regions.
A good example of such a magnetic field was presented by Blandford
\& Znajek (1977); they solved the vacuum Maxwell equations in a 
Kerr space time and derived a split monopole field,
$B^r=B_0 \sin\theta +O(a^2/M^2)$
for a distribution of a toroidal surface current density of 
$I \propto  r^{-2}$.
Specifically, we assume that 
$B^r / (4\pi\mu\eta) = (-E/\mu\sin^2\theta)(1-0.8\cos\theta)^2$
in this paper and calculate $\Ufm$ along a ray path
by solving equation (\ref{eq:def-Ufm}).
In this case, $\vert \eta \vert$ becomes larger in the polar regions;
therefore, the distribution of the fast-surface (Paper 1)

\begin{equation}
  \frac{\rF-\rH}{\rH} 
  = \frac{\pi \Sigma_{\rm H}{}^2}{(\rH-M)M\rH} A_1(\theta) \mu\eta
  \label{eq:fast-surface}
\end{equation}

\noindent
becomes {\it prolate}.
Here, the function $A_1(\theta)$ is of order of unity 
and has a weak dependence on $\theta$ as 

\begin{equation}
  A_1 \equiv \frac{\sqrt{K_{\rm I}-K}}
              {(1-a\OmgH\sin^2\theta)(1-a\OmgF\sin^2\theta)^3W^3},
\end{equation}

\noindent
where $W$ is a function of $\theta$ and of order of unity.
For radial distribution of field lines, $W^2$ becomes

\begin{equation}
  W^2 \equiv 1 +\frac{2a\rH(\OmgH-\OmgF)\sin^2\theta}
             {(\rH-M)(1-a\OmgH\sin^2\theta)(1-a\OmgF\sin^2\theta)}.
\end{equation}

\noindent
The information on the injection point of accretion appears 
only through $K_{\rm I}=0.55$.

Examples of ray paths of axisymmetric waves ($m=0$) are presented in
Fig.\ \ref{figC}.
All the wave packets are radiated from $0 < \theta < \pi/2$ 
(the first quadrant) in this figure.
Even though the wave packets have no angular momenta ($m=0$),
they have non-zero angular velocities 
because of the space-time dragging 
and the rotational motion of the accretion flow.
Therefore, the ray paths are projected on their instantaneous poloidal
plane and are depicted in the figure.
Since $\Ufm$ is smaller in higher latitudes,
the fast-surface becomes prolate;
thus the waves that propagate towards the equator can escape
into the sub-fast regions.
As a result, the wave packets radiated from the first quadrant
propagate clockwise as depicted in this figure.
Because of the accretion, waves are pushed backwards to the hole
and revolve around it.
The heavy solid line on the equatorial plane denotes a dense disk
which possibly resides around an active bole.
If a wave packet collides with the disk, 
it will be totally absorbed to heat the plasma there.

\def\FigC{       
\begin{figure} 
\centerline{ \epsfxsize=8cm \epsfbox[72 230 540 700]{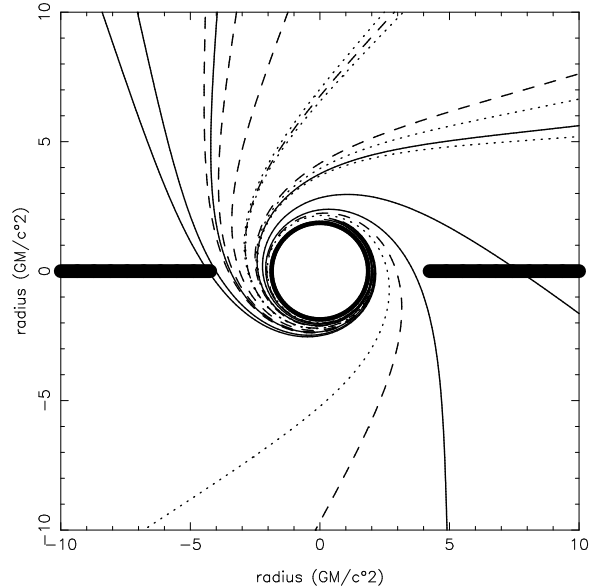} } 
\caption{\label{figC}
  Side view of ray paths of fast-magnetosonic wave packets in a 
  magnetically dominated accretion flow around
  a mildly rotating black hole of $a=0.5M$.
  Ray paths are radiated meridionally from the fast-surface
  every $5$ degrees in the first quadrant:
  Solid curves denote ray paths radiated
  from the high latitudes ($5^\circ$, $10^\circ$, $15^\circ$, 
  $20^\circ$, $25^\circ$, and $30^\circ$). 
  Dashed curves from the middle latitudes 
  ($35^\circ$, $40^\circ$, $45^\circ$, $50^\circ$, $55^\circ$, 
  and $60^\circ$ degrees), while
  dotted curves from the low latitudes 
  ($65^\circ$, $70^\circ$, $75^\circ$, $80^\circ$, and $85^\circ$).
  Fast-magnetosonic speed is lower in the polar regions
  than in the equatorial regions.
  As a result, fast waves collimate towards the rotation axis.
        }
\end{figure} 
}

\FigC
\def\FigD{       
\begin{figure} 
\centerline{ \epsfxsize=8cm \epsfbox[72 230 540 700]{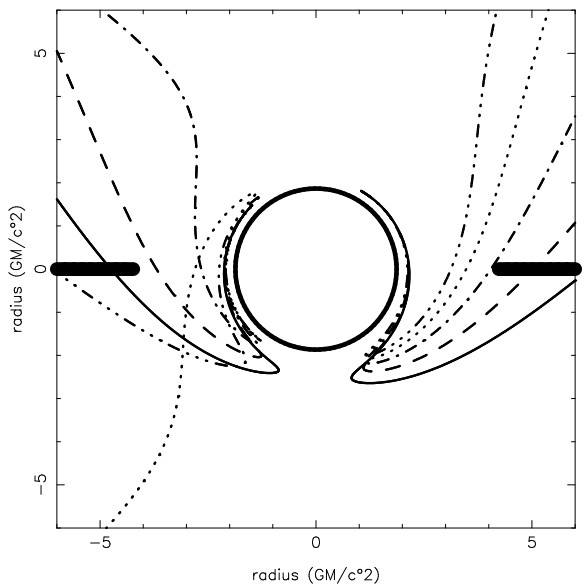} } 
\caption{\label{figD}
  Side view of the ray paths of non-axisymmetric mode.
  The solid lines denote $m=2$ mode, while
  the dashed, dash-dotted, dotted, and dash-dot-dot-dotted lines 
  denote $m=4$, $m=6$, $m=8$ $m=10$ modes, respectively.
  Waves propagating in the right hemisphere are radiated meridionally
  at $\theta=30^\circ$, whereas those in the left one are radiated
  at $\theta=-45^\circ$.
        }
\end{figure} 
}

\FigD
\def\FigE{       
\begin{figure} 
\centerline{ \epsfxsize=8cm \epsfbox[72 230 540 700]{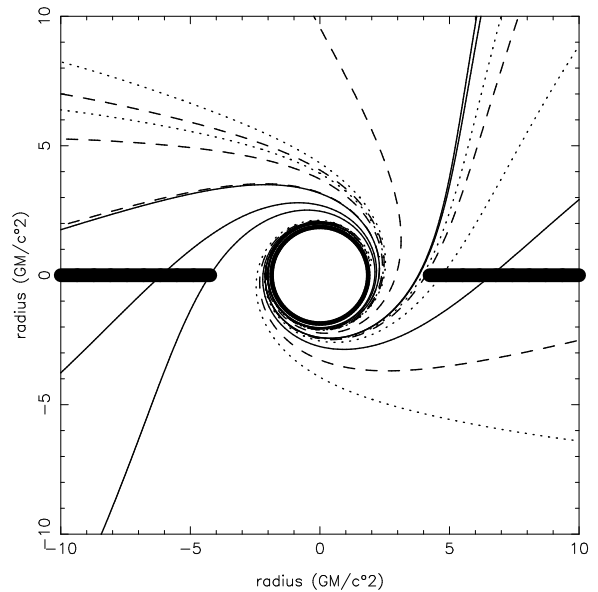} } 
\caption{\label{figE}
  The same figure as Fig. 4; however,
  the fast-magnetosonic speed is slower in the equatorial regions
  than in the polar regions.
  As a result, fast waves are bent towards the equatorial plane
  to form a focal ring.
        }
\end{figure} 
}

\FigE

This figure indicates that {\it most of the wave packets}
which are radiated meridionally from the fast-surface,
{\it collimate into the polar regions where $\Ufm$ is small}.
Qualitatively the same results are obtained for non-axisymmetric
waves.
In Fig.\ \ref{figD}, $m=2, 4, 6, 8, 10$ waves are depicted.
Ray paths in the first and the fourth quadrant originate at 
$\theta=30^\circ$, whereas those in the second and the third
quadrant originate at $\theta=-45^\circ$ (in the second quadrant).
Wave packets with negative angular momentum ($m<0$) are not drawn,
because they are soon swallowed by the hole.
As the figure indicates, 
non-axisymmetric waves cannot reach the rotational axis
because of their non-zero angular momenta;
therefore, the tendency of collimation is 
somewhat weakened compared with $m=0$ modes depicted 
in Fig. \ \ref{figC}.

In section 6, we discuss that the collimated waves may experience 
resonance and mode-convert itself into electromagnetic waves
to result in a particle acceleration by non-linear interactions.
Before we come to this issue, one more point must be clarified:
if $\Ufm$ are larger in the polar regions than 
in the equatorial regions,
ray paths must be bent towards the equatorial plane.
We examine briefly this case in the next subsection. 

\subsection{Formation of a focal ring}

We demonstrate here typical results when the fast-magnetosonic
speed, $\Ufm$, is faster in the polar regions than 
in the equatorial regions.
In this case, the distribution of the fast-surface becomes oblate.
Examples of ray paths for axisymmetric modes ($m=0$) are presented in
Fig.\ \ref{figE}.
All the wave packets are radiated from $0 < \theta < \pi/2$ 
(the first quadrant) in this figure.
Since the fast-surface is oblate, the wave packets that 
propagate initially towards the rotational axis can escape into the
sub-fast regions.
Therefore, the waves propagate counterclockwise.

This figure indicates that most of the wave packets
bent towards the equatorial disk to form a focal ring 
of radius $\sim 5M$ and do not collimate towards the
rotational axis.
For non-axisymmetric waves, this tendency is strengthened 
because of their non-zero angular momenta.

\section{Discussion}

We have demonstrated that the magnetically dominated plasma accretion
becomes highly variable
near to the fast surface located close to the horizon 
and that such fluid's disturbance propagates as 
a fast-magnetosonic wave and collimates towards the rotational axis 
(especially for an axisymmetric mode)
when the fast-magnetosonic speed is slower in the polar regions 
than in the equatorial regions.
In the framework of MHD, the collimated fast waves will not cause 
interesting phenomena such as particle acceleration, 
even in nonlinear regimes.
However, if we take the effects of plasma oscillation and 
cyclotron motion of particles into account, 
interesting results such as particle acceleration at a 
resonant point may be obtained 
(Holcomb and Tajima 1992; Daniel and Tajima 1977).
For this reason, we consider in this section 
a plasma wave in a pure electron-positron plasma 
(for observational and theoretical discussion on the existence of
electron-positron plasmas in AGN jets, see
Ghisellini et al. 1992; Morrison et al. 1992; 
Xie et al. 1995; Reynolds et al. 1996).

As Fig.\ \ref{figC} indicates, 
wave vectors ($\vec{k}$) of collimated fast waves
are nearly parallel to the magnetic field ($\vec{B}$), 
of which poloidal components dominate the toroidal one 
near the rotational axis.
It is, therefore, possible to regard that the collimated waves
propagate as shear Alfv\'en waves.

In fact, a shear Alfv\'en mode is one of the two primary modes
in a pure electron-positron plasma when $\vec{k} \parallel \vec{B}$.
The other mode is an electromagnetic mode.
Their dispersion relation is given by

\begin{equation}
  k^2 = (k_r)^2 
      = \frac{\omega^2(\omega^2-\Omge2-2\omega_{\rm p}{}^2)}
             {c^2(\omega^2-\Omge2)},
  \label{eq:plasma-waves}
\end{equation}

\noindent
where the plasma and cyclotron frequencies are defined by

\begin{eqnarray}
  \omega_{\rm p} &\equiv& \sqrt{4\pi n e^2 / m_{\rm e}},
  \label{eq:def-plasma-freq} \\
  \Omega_{\rm e} &\equiv& \frac{eB^r}{m_{\rm e}c}.
  \label{eq:def-cyclo-freq}
\end{eqnarray}

\noindent
If we would set $\Omega_{\rm e} \gg \omega,\ \omega_{\rm p}$,
we could recover the dispersion relation of an usual Alfv\'en mode,
which is identical with that of a fast-magnetosonic mode when 
$\vec{k} \parallel \vec{B}$ in a cold plasma.
Exactly speaking, we must take account of geometrical correction
due to hole's gravity in the definitions 
(\ref{eq:def-plasma-freq}) and (\ref{eq:def-cyclo-freq});
however, at the height where $r>10M$, such effects become negligible.

From the dispersion relation (\ref{eq:plasma-waves}),
it follows that a shear Alfv\'en mode exists for frequencies
less than $\Omega_{\rm e}$ (the resonance frequency),
while an electromagnetic mode exists for frequencies greater than
$\sqrt{\Omge2 + 2\omega_{\rm P}}$ (the cut-off frequency).
In a realistic black hole magnetosphere,
$\Omega_{\rm e}$ is a decreasing functions of distance ($r$)
from the hole, because $B^r$ decreases with $r$.

It is important to note that the point of cut-off is located 
slightly outside of the point of resonance in a magnetically
dominated magnetosphere 
($m_{\rm e} n_{\rm e} c^2 \ll B^2 / 8\pi$).
Therefore, we can depict the following scenario according to 
\cite{bud61}:
A shear Alfv\'en wave packet is injected outwards by the mechanism
described in the preceding sections.
As the wave approaches the point of resonance (a magnetic beach),
it increases the amplitude owing to a nonlinear effect.
After reaching the point of resonance,
it evanesces through the thin evanescent region 
to transmit as an electromagnetic wave above the point of cut-off.

When the amplitude of injected shear Alfv\'en wave is high and
nonlinear effects become important,
the wave may capture many particles to hide the singularity.
In this case, they continue to propagate as an Alfv\'en mode
after passing the \lq singularity' owing to the so-called
self-induced tunneling (e.g., McCall, Hahn 1969).
As a result, the trapped particles may be accelerated through
long distances to become relativistic.
According to the nonlinear simulations performed by
Daniel and Tajima (1977),
who adopted a very thin evanescent layer corresponding to
a magnetic dominance of $B^2/8\pi \sim 9n_{\rm e}m_{\rm e}c^2$,
particle acceleration up to energies of $8m_{\rm e}c^2$
can be realized if the injected wave is highly nonlinear.
On these grounds, 
it seems reasonable to conclude that the process of wave 
amplification and collimated demonstrated in this paper
triggers initial acceleration of jet 
due to a cyclotron resonance in an electron-positron plasma.

\acknowledgments  

Thanks are due to Drs. A. Tomimatsu, K. Shibata, and M. Takahashi
for valuable advice and helpful suggestions. 

\clearpage

%
%

\clearpage

\end{document}